\documentclass[preprint,11pt,authoryear]{elsarticle}

\usepackage{amssymb}
\usepackage{lipsum}
\usepackage{amsmath}

\usepackage{tensor}
\usepackage{xspace} 
\usepackage{tabularx}
\usepackage{multirow}
\usepackage{dsfont}
\usepackage{natbib}

\usepackage[draft=false]{hyperref}

\hypersetup{colorlinks,linkcolor={red},citecolor={blue},urlcolor={blue}}  

\usepackage[legalpaper, margin=2cm]{geometry}

\journal{Physics Letters B}

\begin{document}

\begin{frontmatter}



\title{Constraints on non-local gravity from binary pulsars gravitational emission}


\author[third]{Amodio Carleo}



\affiliation[third]{organization={INFN, sezione di Napoli, Gruppo collegato di Salerno},
addressline={Via Giovanni Paolo II, 132}, 
city={Fisciano (SA)},
postcode={I-84084}, 
country={Italy}}

\begin{abstract}
Non-local theories of gravity are considered extended theories of gravity, meaning that when
the non-local terms are canceled out, the limit of General Relativity (GR) is obtained. Several reasons have led us to consider this theory with increasing interest, but primarily non-locality  emerges in a natural way as a ’side’ effect of the introduction of quantum corrections to GR, the purpose of which was to cure the singularity problem, both at astrophysical and cosmological level. In this paper we studied  a peculiar case of the so called Deser-Woodard  theory  consisting in the addition of a non-local term of the form $R \Box^{-1}R$ to the Hilbert-Einstein lagrangian,
where $R$ is the Ricci scalar, and derived, for the first time, contraints on the dimensionaless non-local parameter $A$ by exploiting the predicted gravitational wave emission in three binary pulsars, namely PSR J1012+5307, PSR J0348+0432 and PSR J1738+0333. We discovered  that the instantaneous flux strongly depends on $A$ and that the best constraints ($0.12 < A < 0.16$) come from PSR J1012+5307, for which the GR prediction  is outside the observational ranges. However, since for PSR J$1012 + 5307$ scintillation is suspected, as emerged in a recent census by LOFAR  corruptions in pulsar timing could be hidden. We finally comment on the usability and reliability of this type of test for extended theories of gravity.

\end{abstract}




\end{frontmatter}




\section{Introduction}
\label{introduction}

Einstein’s local principle of equivalence  is based on the
assumption that an accelerated observer in Minkowski spacetime, at each
event along its world line, is physically equivalent to a momentarily identical
comoving inertial observer. When the past history of the accelerated observer is also considered, non-locality, a  typically quantum property,  comes out. 
It is not a new fact that Quantum Mechanics shows non-local aspects: non-locality is a manifestation of entanglement and the latter is has been repeatedly demonstrated in laboratory experiments; the Bell's theorem, furthermore, demonstrated that locality is violated in some quantum systems. After that, non-locality  has been investigated in the field of QFTs (see e.g. \cite{Pauli:1949,Pauli:1953,Efimov:1,Efimov:2,Efimov:3,Efimov:4,Efimov:5}). 
Gravitational theories with this feature   have recently
been developed  with the aim of combine the quantum world with General Relativity (GR), while addressing,  at the same time, some of the most important issues of modern Cosmology, namely dark energy or dark matter. Moreover,  in string theory  \citep{String_1,string_4} non-locality mainly  emerges as a 'side' effect of the introduction of quantum corrections to  (GR) the purpose of which was to cure the singularity problem.  Indeed, applications to Cosmology showed that non-local ghost-free higher-derivative
modifications of Einstein gravity in the ultraviolet regime can  admit non-singular bouncing solutions for the Universe (in place of the Big Bang singular  solution) and non-singular Schwarzshild metrics for black-holes \citep{Modesto_2011}. Another motivation for considering non-local gravity is the possibility to achieve renormalizability without the appearance of ghost modes \citep{Modesto_Renormalizab}.  
A road towards Quantum Gravity by  considering non-local corrections to the Hilbert-Einstein action has been drawn in  \citep{Modesto:2015lna,Modesto:2015ozb}. The introduction of non-local terms
also appeared  in alternative theories of gravity, such as teleparallel gravity \citep{Bahamonde:2017bps}. However, it must be said that  the non-local theories of gravity  themselves are considered extended theories of gravity, meaning  that when the non-local terms are canceled out, the limit of the GR is obtained. More precisely, non-local theories of gravity are described by Lagrangians composed of a finite sum of products between fields and
their derivatives evaluated at different points $x$ and $y$
of the spacetime while metric $g_{\mu\nu}$ and/or other fields  are
described by integro-differential equations, implying that the value of the field at one point depends on its
value at another point of the spacetime, weighted by a function called nucleus or kernel. \\

There are essentially three different ways to implement non-locality from a mathematical point of view. The first manner (the most studied in the literature)
is by means of a convergent series expansion with real coefficients $c_n$ of an analytic non-polynomial
function $F$ of the D'Alembert operator $\Box$, known as Infinite Derivative of Gravity \citep{Efimov:1,Buoninfante_1,Buoninfante_2,Tesi_IDG_Edholm}. Recently, it was shown that in gravity theories containing such   class of non-local terms 
the linearized Ricci tensor $R_{\mu\nu}$ and  Ricci scalar $R$ are not vanishing in the region of non-locality, i.e. at short distance from a source, due to the smearing of the source induced by the presence
of non-local gravitational interactions. It follows that, unlike in Einstein’s gravity, the
Riemann tensor is not traceless and it does not coincide with the Weyl tensor, which, however, vanishes
at short distances, implying that the (static) metric is  conformally flat in that region
\citep{Buoninfante_1}, implying a possible deviation from $1/r$ potential drop at very short distance \footnote{A decay as $1/r$ of the gravitational potential has only been verified up to $\sim 10^{-5}$ m, which is thirty orders of magnitude away from the Planck length \citep{bilancia_tors}.}. In the second way, that we will follow, the non-locality  manifests itself in non-analytic operators such as  $\Box^{-n}$.  It was shown that the application of the non-local operator $\Box^{-1}$
to the scalar curvature $R$ gives rise to the late-time cosmic expansion of the Universe without invoking any Dark Energy contribution. For an overview on non-local cosmology see also \cite{Review_Non_local}. Finally, non-locality can enter  through a constitutive relation on the (linearized) gravitational field involving a causal
kernel  determined via observational data, in the spirit of non-local electrodynamics of media \citep{Mashhoon:2022,Puetzfeld_2019}.

In all these approaches, it is important to study the linearized versions of the theories and to
derive gravitational waves (GWs). For IDG, they have been studied in \cite{GWs_IDG_2012} and \cite{GWs-IDG_2018}, while for higher order theories with lagrangian  $\mathcal{L}= R + \sum_{h=1}^{n} a_h R \Box^{-h} R$ with $n$ fixed they are given in \cite{Capriolo}. Indeed, gravitational radiation allows to detect possible effects
of non-local gravity \citep{Capcap:2020xem} as well as to classify the degrees of freedom of a given theory. Here, we exploit the dynamics of well-known binary systems to constraint the free parameter(s) of the theory. In particular, we used astrophysical data for three binary pulsar, namely PSR J1012+5307, PSR J0348+0432 and PSR J1738+0333.  We consider as a test tool the derivative of the orbital period and its variation \citep{Stairs_2003}, which is one of the best estimated parameters.  In phenomenological software like \texttt{TEMPO} or \texttt{TEMPO2}, it is one of the output fitting parameters and it obtained without assuming any theoretical framework. This is a crucial point in all the cases in which the gravity theory is supposed not to be the GR. For this reason, these types of tests are quite common in extended or alternative theories of gravity; see for example \cite{Laurentis_2011,Laurentis_2013,0333,Nazari_2022}.

\section{The non-local model}

In this section we derive the field equations for an extended theory of gravity given by the action \cite{2107.06972} 
\begin{equation}\label{action}
 S[g] = \dfrac{1}{2 \chi} \int d^4 x \sqrt{-g} \Big( R + A R \Box^{-1} R  \Big) + \int d^4 x \sqrt{-g} \mathcal{L}_m[g]
\end{equation}
where $A$ is an dimensionless constant,  $\chi= 8 \pi G$, $\mathcal{L}_m$ is the matter lagrangian, $\Box^{-1}$ is the inverse operator of the D'Alembert one $\Box = \nabla^{\mu}\nabla_{\mu}$, being $\Box^{-1} \Box = \mathds{1}$ . The action above is the simplest case of the Deser-Woodard gravity theory \citep{DW_original} and, more precisely, it is one of the only two possible models allowed by the Noether symmetry approach \citep{S2}, the other one  being  characterized by a non-local correction $\sim R  \exp \{A \Box^{-1}R\}$. It is clear from Eq. (\ref{action}) that, due to  the non-local term $R\Box^{-1}R$, the field equations are very involved and non-linear. One way to bypass this problem is to build a localized version of (\ref{action}), by introducing the auxiliary field $\phi = \Box^{-1} R$. The variation with respect to the metric $g^{\mu\nu}$  gives the main equation of motion, i.e.
\begin{equation}\label{eq:fe}
    G_{\mu\nu} + (G_{\mu\nu} + g_{\mu\nu} \Box - \nabla_{\mu} \nabla_{\nu}  ) (a\phi - \lambda ) - \nabla_{(\mu}\phi \nabla_{\nu)}\lambda + \dfrac{1}{2} g_{\mu\nu} \nabla^{\sigma}\phi\nabla_{\sigma}\lambda = \chi T_{\mu\nu} ,
\end{equation}
while variations with respect to both the scalar fields give the two constraints 
\begin{equation}\label{constr}
    \Box \lambda = -A R \;, \; \; \; \; \Box \phi =  R 
\end{equation}
where $\lambda$ is the Lagrange multiplier. After a long but standard procedure, the gravitational wave stress-energy tensor (GW SET) is found to be  

\begin{equation}\label{pseudo_final}
    \tau^{\alpha}_{\beta} = \dfrac{\psi_0}{32 \pi G } \Big\langle   \partial^{\alpha} \theta^{\mu\nu}_{(TT)}  \partial_{\beta} \theta_{\mu\nu}^{(TT)} + \dfrac{6}{\psi_0^2} \partial^{\alpha}\psi^{(1)} \partial_{\beta} \psi^{(1)} + 
    \dfrac{4}{\psi_0} \partial^{\alpha}\varphi \partial_{\beta} w \Big\rangle 
\end{equation}

where we introduced the new fields

\begin{equation}\label{gauge}
    \theta_{\mu\nu} \doteq  h_{\mu\nu} - \dfrac{1}{2} \eta_{\mu\nu} h - \dfrac{\eta_{\mu\nu}}{\psi_{0}}\psi^{(1)} \; , \; \; \; \psi^{(1)}= A \varphi - w .
\end{equation}

Above, we have stressed the gauge condition on $\theta_{\mu\nu}$ and brackets means average over all wavelenghts. Notice that the expression above is symmetric in indices  $\alpha$, $\beta$ and gauge-invariant only under average procedure; in this sense, it is called pseudotensor. In order to get physical quantities from the (pseudo)tensor (\ref{pseudo_final}), we have to link each of the fields $\theta_{\mu\nu}, \varphi, w$ to the source of gravitational waves, that is to $T_{\mu\nu}$. This implies to solve the equations of motions not in vacuum, but with $T_{\mu\nu}\not=0$. The field equations with source and in the Lorenz gauge  are \footnote{Since our background space is Minkowski, we have $\phi_{0}= \Box^{-1} R^{(0)} = 0 $ and, similarly, $\lambda_{0}=0$. Furthermore, notice that in solving non-vacuum equations the traceless condition is not allowed.   } 

\begin{equation}\label{field_eqs}
    \Box \theta_{\mu\nu} = -2 \chi  T_{\mu\nu} \; , \; \; \; \Box \varphi = \dfrac{1}{2} \Box \theta + 3 \Box \psi^{(1)}   \; , \; \; \; \Box w = - A \left[ \dfrac{1}{2} \Box \theta + 3 \Box \psi^{(1)} \right].
\end{equation}

After excluding the case $A=1/6$ (it would imply a traceless condition  on $T_{\mu\nu}$, which is not physically reasonable for binary systems), the only solution of the above system of equations is the following: 

\begin{equation}\label{sol_2}
A\not=1/6 \; \; \; \; \; \; \; \Rightarrow \; \; \; \; \; \; \;  \Bigl\{ \Box \varphi = \dfrac{1}{6A-1} \chi T \; , \; \Box w = \dfrac{-A}{6A-1} \chi T   \Bigr\}
\end{equation}
which, once solved,  clearly give explicit solutions for the scalar fields $\varphi$, $w$ in terms of the source. 
The most important physical quantity obtainable from the pseudotensor is the energy flux. More precisely,  in the far region condition, the well-known quadrupolar approximation is used,  and, using the field equations (\ref{field_eqs}), the total emitted power (or luminosity) is given by \citep{Maggiore:I}
\begin{equation}\label{P_tot}
    P_{tot} = - r^2 \int d\Omega \langle \tau^{0 i} \rangle  n_i := P_{tot}^{GR} + P_{tot}^{NL}
\end{equation}
with integration on a spatial surface at spatial infinity. The subscript "NL" labels the non-local contribution in addition to the GR result, while $n_i \doteq \hat{x}_i = x_i/r$, with $r$ the distance to the binary system.  In the formalism of the quadrupole tensor, $ Q_{ij}(t) \doteq \int d^3 y T^{00}(t,\mathbf{y}) y^i y^j$, and after standard computations Eq.(\ref{P_tot}) gives 

\begin{equation}\label{p_tot_finale}
 P_{tot} = \dfrac{G}{5 c^5} \Big[ \Big(1+ \dfrac{A}{3 (6A-1)} \Big) \langle \dddot{Q}^{ij}\dddot{Q}_{ij} \rangle + \dfrac{1+7A}{3(6A-1)} \langle \dddot{Q}^2 \rangle   \Big]  \; .
\end{equation}
When the non-local parameter $A$ is zero, then the usual GR result is recovered \citep{Maggiore:I}. The  result above shows that non-local corrections are, at least in principle, compatible with orbit decays by quadrupole  radiation in binary systems, paving the way for new observational constraints on non-local corrections compared to those already present in the literature \citep{Amendola_2019, S2}. Of course, to find constraints on the non-local parameter $a$ we need to apply Eq. (\ref{p_tot_finale})  to real astrophysical systems. Indeed,  to our knowledge, there are no limits on the parameter $A$ until today. This way to proceed, i.e. constraining parameters after the selection of  the functional form for the correction, is alternative to a more phenomenological one, where the functional form itself can be selected by fitting with the observations. 

\section{Constraints from pulsars}
\label{sec4}
In this section we apply Eq.(\ref{p_tot_finale}) to three well-known binary pulsars, namely PSR J1012+5307, PSR J0348+0432 and PSR J1738+0333, whose main observed data (distance, observed orbital period derivative  and masses) are reported in Table \ref{Table1}. We called $m_p$ the mass of the pulsar and $m_c$ the mass of the companion star; with $\Dot{P}_b$ we indicated the observed variation of the period as phenomenologically estimated  by  \texttt{TEMPO}\footnote{The value $\Dot{P}_b$ provided by the \texttt{TEMPO} fitting   does not take into account some kinematic effects due to the relative motion between the binary system and the center of gravity of the solar system. In order to make a comparison with the analytical results of this section, it is necessary to correct the values by subtracting such effects. The corrected value was called $\Dot{P}_b^{corr}$.} (see caption of Table \ref{Table1} for references); finally, we also listed the predicted value according to General Relativity $\Dot{P}_b^{GR}$ and the corrected observed value $\Dot{P}_b^{corr}$ given by the formula \citep{5307C}

\begin{equation}
\Dot{P}_b^{corr} =     \Dot{P}_b - \Dot{P}_{Shk} - \Dot{P}_{Gal} - \Dot{P}_b^{\Dot{m}} - \Dot{P}_b^{T}
\end{equation}
In the relation above, the first two corrections to the  \texttt{TEMPO} value $\Dot{P}_b$  are due to the 
transverse motion of the source and to the difference in the galactic acceleration, respectively, and  are given by  
\begin{equation}
  \Dot{P}_{Shk}=   \dfrac{P_b a_s}{c} \;, \; \; \;  \Dot{P}_{Gal}=   \dfrac{P_b a_{Gal}}{c} 
\end{equation}
where $P_b$ is the binary period of the system, $a_s$ is  the apparent acceleration from the Shklovskii effect\footnote{Among all the corrections, the Shklovskii effect is the most relevant one. It depends on the distance $d$ of the source and its total proper motion $\mu_T$. If the orbital period $P_b$ is expressed in seconds and the the speed of light in $m/s$, then the acceleration $a_s$, in $m/s^2$, is given by $a_s = 0.7271 \times 10^{-12} (\frac{d}{kpc}) (\frac{\mu_T}{mas \; yr^{-1}})^2 $ (see Eq. (16) in \cite{accelerations_effects}). } \citep{Shklovskii_1970} and $a_{gal}$ is galactic acceleration at the position of the binary system \citep{accelerations_effects}.  The terms $\Dot{P}_b^{\Dot{m}}$  and $\Dot{P}_b^{T}$ 
are contributions intrinsic to the binary system resulting from a mass loss or a deformation in the companion star, respectively. These last two contributions are usually negligible and for this reason we will not consider them. Notice that in some other references the term $\Dot{P}_b^{corr}$ is named $\Dot{P}_b^{GW}$, since it represents the orbital period variation due only to the gravitational emission. \\

As is usual in the treatment of binary systems, we use the reference frame of the center of mass (CM), in which the two-body problem is reduced to a one-body problem, namely a particle of mass $\mu=(m_p m_c)/m$, with $m=m_c+m_p$ the total mass, in orbit on an elliptical trajectory and subject to an acceleration $\mathbf{\Ddot{r}}= - (G m /r^2)\mathbf{\hat{r}}$ where $\mathbf{r}$ is its position to the origin. The modulus $r$ is given by 

\begin{equation}
    r = \dfrac{a (1-e^2)}{1+e \cos \psi} \; , \; \; \; a = \dfrac{G M \mu }{|E|}
\end{equation}

where $a$ is the semi-major axis,$e$ the eccentricity, $\psi$ the true anomaly and $E$ the gravitational energy of the system. In this reference frame, the mass quadrupole of $\mu$ is equal to the mass quadrupole of the iniatial two masses. From \cite{Maggiore:I} we then obtain in our notations

\[\arraycolsep=1.4pt\def\arraystretch{1.6}
\begin{array}{l}
\dddot{Q}_{11}=\beta(1+e \cos \psi)^2\left[2 \sin 2 \psi+3 e \sin \psi \cos ^2 \psi\right] \\
\dddot{Q}_{22}=\beta(1+e \cos \psi)^2\left[-2 \sin 2 \psi-e \sin \psi\left(1+3 \cos ^2 \psi\right)\right], \\
\dddot{Q}_{12}=\beta(1+e \cos \psi)^2\left[-2 \cos 2 \psi+e \cos \psi\left(1-3 \cos ^2 \psi\right)\right],
\end{array}
\]

where $\beta = \dfrac{4 G^3 \mu^2 m^3}{a^5 (1-e^2)^5}$. Therefore, the GR contribution in Eq. (\ref{p_tot_finale}) reads as 

\begin{equation}
    P_{tot}^{GR}(\psi) = \dfrac{8 G^4 \mu^2 m^3}{15 c^5 a^5 (1-e^2)^5} \Big\langle (1+e \cos \psi)^4 [e^2 \sin^2 \psi + 12 (1+e \cos \psi)^2] \Big\rangle
\end{equation}

while the non-local contribution is given by 

\begin{equation}
    P_{tot}^{NL}(\psi) = \dfrac{4 A \mu^2 m^3 G^4}{15 c^5 a^5 (6A-1)(1-e^2)^5}\Big\langle (1+e \cos \psi)^4 [8+11 e^2 + 16 e \cos \psi - 3 e^2 \cos (2 \psi)]  \Big\rangle .
\end{equation}

Putting all together and after using the relation \citep{Maggiore:I}

\begin{equation}
    \Dot{\psi} = \sqrt{\dfrac{G m}{a^3}} (1-e^2)^{-3/2}(1+e \cos \psi)^2 
\end{equation}
we found the total power 

\begin{equation}
    P_{tot} = \dfrac{(1-e^2)^{3/2}}{2 \pi}\int_{0}^{2 \pi} d\psi (1+e \cos \psi)^{-2} \big[ P_{tot}^{GR}(\psi)+P_{tot}^{NL}(\psi)\big] = \dfrac{32 G^4 m^2 \mu^2}{5 c^5 a^5} f(e,A)
\end{equation}
where we defined the function 

\begin{equation}
    f(e,A)= \dfrac{e^4(241 A-37)+4 e^2 (469 A -73) + 608 A-96}{96(6A-1)(1-e^2)^{7/2}}.
\end{equation}

When $A=0$ it is easy to verify that the above function is reduced to the well-known enhancement factor of GR \citep{Maggiore:I}. \\
In order to compute the theoretical estimation for the variation of the orbital period, $\Dot{T}$, we notice that $\frac{\Dot{T}}{T} = - \frac{3 \Dot{E}}{2 E}$, where $T=2\pi \sqrt{\frac{a^3}{G m}}$ is the period, $\Dot{E}=-P_{tot}$, while $E=-\frac{G m \mu}{2a}$. Since $a$ can be written also as $a= G^{1/3} m^{1/3} (T/(2\pi))^{2/3}$, we arrive at 

\begin{equation}\label{T_dot}
    \Dot{T}= - \dfrac{192}{5} \dfrac{G^{5/3}m^{2/3}\mu \pi}{c^5} \Big(\dfrac{T}{2 \pi}\Big)^{-5/3} f(e,A).
\end{equation}

Here, a comment is in order. The above relation, as a function of $A$, has to be compared with the correct observed value  $\Dot{P}_b^{corr}$ discussed before.  In this procedure, one can use without problems all the parameters fitted by \texttt{TEMPO} except for the masses, as the values of these require a gravity theory a priori \citep{weisberg_1989}. Therefore, wanting to use the variation of the orbital period to set constraints on alternative theories of gravity, one would have to look for estimates of the masses in an independent way. This is a very crucial point, almost never underlined in the literature. One way to get such masses is to use two additional post-keplerian parameters (such as $\Dot{w}$ and $\gamma$), obtained in the specific gravity theory. In our case, they would depend on the non-local parameter $A$, just like $\Dot{T}$. However, this approach could be very hard-working and nullified by the fact that  $\Dot{T}$ is the best measured post-keplerian parameter. An alternative, in our opinion, could be offered by pulsar-white dwarf  binary systems, as in some of these systems the mass of the companion star   is obtained from optical observations. From the measurement  of the radial velocity of the two bodies, the mass ratio $q=m_p/m_c$ is estimated, thus allowing measurements to be obtained for both masses, without ever exploiting timing measurements. In Table \ref{Table1}, all sources are of this type.

\begin{center}
\begin{table}
\setlength{\tabcolsep}{5pt} 
\renewcommand{\arraystretch}{1.5} 
\begin{tabular}{c c c c c c c c } 
 \hline
 
 Source & Distance  & $\Dot{P}_b$ & $m_p$   & $m_c$ & $\Dot{P}_b^{GR}$ & $\Dot{P}_b^{corr}$ \\ 
        &    (kpc)   &   ($\times 10^{-14}$)         &  ($M_{\odot}$)  &   ($M_{\odot}$) & ($\times 10^{-14}$)   &   ($\times 10^{-14}$)   \\
 \hline
 
PSR J1012+5307 & 	$0.828^{+0.056}_{-0.018}$  & $6.1\pm0.4$ & $1.72\pm0.16$ & $0.165\pm0.015$ & $-1.16\pm0.19$ & $-0.2\pm 0.6$\\ 

 PSR J0348+0432 & $2.1 \pm 0.2$ & $-27.3 \pm 4.5$ & $2.01 \pm 0.04$ & $0.172 \pm 0.003$ & $-25.8^{+0.8}_{-1.1}$ & $-27 \pm 5$	 \\ 

 PSR J1738+0333 & $1.47 \pm 0.10$ & $-1.70 \pm 0.31$ & $1.46 \pm 0.06$ & $0.181 \pm 0.008$ & $-2.77^{+0.15}_{-0.19}$ & $-2.59 \pm 0.32$	 \\ 
 \hline
\end{tabular}
\caption{List and main data for the three pulsar used in our analysis. In particular, $m_p$ is mass of the pulsar; $m_c$ the mass of the companion star; with $\Dot{P}_b$ we indicated the observed variation of the period as phenomenologically estimated  by  \texttt{TEMPO}; finally, $\Dot{P}_b^{GR}$ is the  predicted value according to General Relativity  and $\Dot{P}_b^{corr}$ the corrected observed value after subtracting the kinematic effects from $\Dot{P}_b$. For this table and the entire manuscript, we used as references \citep{5307A,5307B,5307C} for  PSR J$1012+5307$, \citep{0432A,0432B} for PSR J$0348+0432$, and \citep{0333A} for PSR J$1738+0333$. Notice that in literature there are different estimates for $\Dot{P}_b^{GR}$ and $\Dot{P}_b^{corr}$ essentially due to different estimates in the distance or in the star masses. For this reason, our estimates for PSR J$1012+5307$ are different from \cite{5307C}. All uncertainties are at 1$\sigma$. 
}

\label{Table1}
\end{table}
\end{center}

\begin{center}
\begin{table}
\setlength{\tabcolsep}{10pt} 
\renewcommand{\arraystretch}{1.5} 
\begin{tabular}{c c c c  } 
 \hline
 
 Source & $P_b$  & $\Dot{P}_{Shk}$ &  $\Dot{P}_{Gal}$  \\ 
        &    (day)   &  ($\times 10^{-14}$)  &  ($\times 10^{-14}$)    \\
 \hline
 
PSR J1012+5307 & 	$0.60467271355(3)$  & $6.89^{+0.50}_{-0.15}$ & $-0.55\pm0.02$ \\ 

 PSR J0348+0432 & $0.102424062722(7)$ & $0.129^{+0.025}_{-0.021}$ & $0.037^{+0.006}_{-0.005}$ 	 \\ 

 PSR J1738+0333 & $0.3547907398724(13)$ & $0.83^{+0.06}_{-0.05}$ & $0.058^{+0.016}_{-0.014}$	 \\ 
 \hline
\end{tabular}
\caption{ Orbital period  $P_b$ for the three binary pulsars, as estimated by \texttt{TEMPO}, and the corrections to $\Dot{P}_b$ due to the Shklovskii effect, $\Dot{P}_{Shk}$, and the galactic correction, $\Dot{P}_{Gal}$ (see caption in Table \ref{Table1} for references). Digit(s) in parentheses are uncertainties in the last digit quoted, as
estimated by \texttt{TEMPO}. All uncertainties are at 1$\sigma$. 
}
\label{Table2}
\end{table}
\end{center}

\section{Results}
\begin{figure}
	\centering 
	\includegraphics[width=1.0\textwidth]{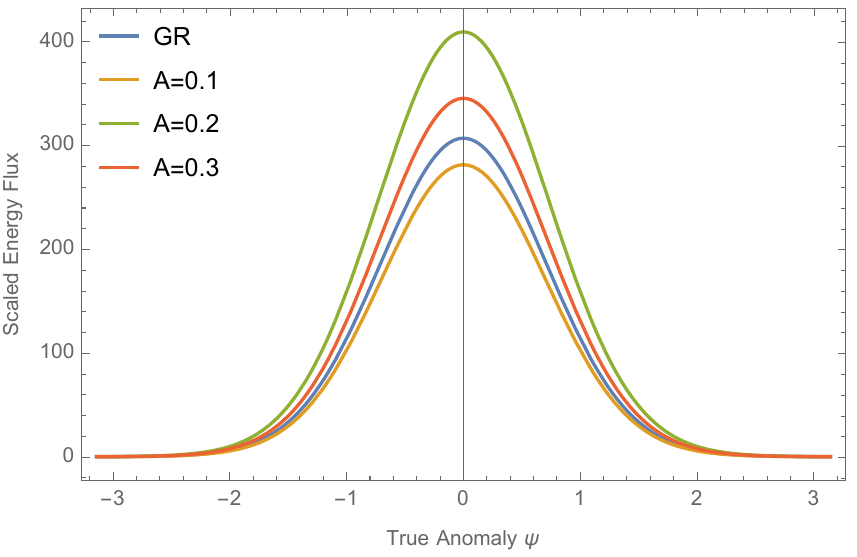}	
	\caption{dimensionless instantaneous flux of gravitational energy as function of true anomaly $\psi$ for different values of $A$.  It is clear the strong dependence on the non-local parameter $A$: when $A<1/6$ the flux is weaker than the GR equivalent, but it increases as $A$ decreases;  when $A>1/6$ the flux is stronger than the GR equivalent, but it decreases as $A$ increases. The most probable range for $A$, i.e. the range of values that deviate the least from the GR prediction,  is $A \ll 1$, as expected. An eccentricity value $e=0.5$ has been set.} 
	\label{fig1}
\end{figure}

\begin{figure}
	\centering 
	\includegraphics[width=1.0\textwidth]{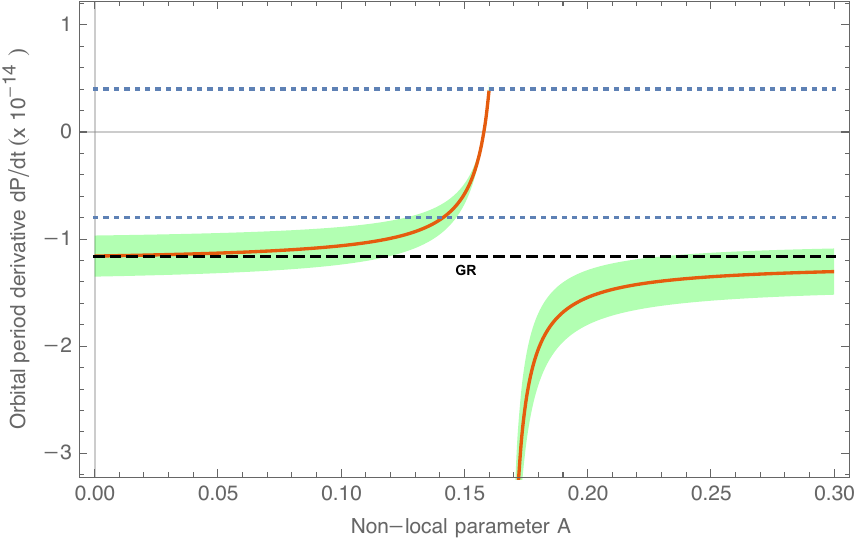}	
	\caption{Plot of $\Dot{T}$ as a function of $A$ for PSR J$1012+5307$.  The range of the observed value ($\Dot{P}_b^{corr}$) is between the two blue dotted lines: if the theoretical prediction crosses this region then the theory is admissible. While the GR line (dotted black) lies largely outside the  region (even considering the error bands), the non-local prediction (orange line) with the corresponding error band (green shading) crosses the strip with non-zero $A$ values. Notice that when $A=0$ the non-local curve reaches the GR prediction, as expected and that the feasible range for $A$ seems to be approximately $0.12 < A < 0.16$. }  
	\label{fig2}
\end{figure}

We corrected $\Dot{P}_b$ by subtracting both the Shklovskii effect and the galactic correction, whose values are listed in Table \ref{Table2}, along with the orbital period  derived by \texttt{TEMPO}, $\Dot{P}_b$. Notice that the latter is used to compute the kinematic corrections and to estimate $T$ in Eq. (\ref{T_dot}). Using the data in Tables \ref{Table1} and \ref{Table2}, the r.h.s. of Eq. (\ref{T_dot}) provides a numerical value with a corresponding  uncertainty, to compare with the corrected observed value $\Dot{P}_b^{corr}$. In Fig. (\ref{fig1}) we first computed the instantaneous flux of gravitational energy as function of true anomaly $\psi$ for different values of $A$. More precisely, we plotted the function $P(\psi) = P_{tot}^{GR}(\psi)+P_{tot}^{NL}(\psi) $, divided by the factor $\frac{m^3 \mu^2 G^4}{c^5 a^5}$, in order to obtain an dimensionless quantity. It is clear the strong dependence on the non-local parameter $A$: when $A<1/6$ the flux is weaker than the GR equivalent, but it increases as $A$ decreases;  when $A>1/6$ the flux is stronger than the GR equivalent, but it decreases as $A$ increases. Therefore, based on this, the most probable range for $A$, i.e. the range of values that deviate the least from the GR prediction,  is $A \ll 1$, as expected. In Fig. (\ref{fig2}) the plot of $\Dot{T}$ as a function of $A$ for PSR J$1012+5307$ is shown. We chose this source since it is the only  one in our list that disagrees with the GR prediction (black dotted line). The range of the observed values is between the two blue dotted lines: if the theoretical prediction crosses this region then the theory is admissible. While the GR line  lies largely outside the  region, the non-local prediction (orange line) with the corresponding error band (green shading) crosses the strip with non-zero $A$ values. The error bands for the theoretical prediction  are due to the uncertainties in four quantities\footnote{The uncertainty on the distance of the source, which, together with those on the masses, is the most significant, however, affects  the range of the observed value, as the kinematic corrections strongly depend on the distance.}, all estimated by \texttt{TEMPO}: $m_c$, $m_p$, $P_b$ and $e$.  Notice that when $A=0$ the non-local curve reaches the GR prediction, as expected. From this figure, it seems that there is room for a non-local contribution in this binary system, at least for the post-Keplerian parameter we considered and  that the feasible range for $A$ is approximately $0.12 < A < 0.16$, which is, to our knowledge,  the first constraint on the non-local parameter $A$.

\section{Summary and conclusions}

The use of binary pulsars was the first and most widely used system for testing both GR and modified theories of gravity. Although in the literature the binary system most used as an instrument is PSR $1913+16$, this source is not suitable for such a purpose if only one post-keplerian parameter is studied. Indeed, in this case, one is tempted to use the mass values reported in literature, forgetting that to obtain those values other post-Keplerian parameters were used, usually  with the GR assumed to be true \citep{weisberg_1989}. This mixture of underlying theories cannot be useful for making predictions about the theory under study. Therefore, wanting to use the variation of the orbital period to set constraints on alternative theories of gravity, one would have to look for estimates of the masses in an independent way. This is a very crucial point, almost never underlined in the literature. One way to get such masses is to use two additional post-keplerian parameters (such as $\Dot{w}$ and $\gamma$), obtained in the specific gravity theory. In our case, they would depend on the non-local parameter $A$, just like $\Dot{T}$. However, this approach could be very hard-working and nullified by the fact that  post-keplerian parameters different from $\Dot{T}$ are not measured with great accuracy. An alternative strategy, in our opinion, could be offered by pulsar-white dwarf  binary systems, as in some of these systems the mass of the companion star   is obtained from optical observations. From the measurement  of the radial velocities of the two bodies, the mass ratio $q=m_p/m_c$ is also estimated, thus allowing measurements to be obtained for both masses, without ever exploiting timing measurements. 

In this paper we followed the second approach, choosing a single post-Keplerian parameter (the variation of the orbital period, $\Dot{P}_b$) and therefore using three binary pulsars having a white-dwarf as companion star, with all masses obtained spectroscopically or astronometrically. All data are listed in Tables \ref{Table1} and \ref{Table2}. We computed the total emitted power (or luminosity) as a function of the true anomaly and the theoretical prediction for orbital period variation, $\Dot{T}$, emphasizing the non-local contribution. We found that the instantaneous flux of gravitational energy  strongly depends  on the non-local parameter $A$: when $A<1/6$ the flux is weaker than the GR equivalent, but it increases as $A$ decreases;  when $A>1/6$ the flux is stronger than the GR equivalent, but it decreases as $A$ increases. Therefore, based on this, the most probable range for $A$, i.e. the range of values that deviate the least from the GR prediction,  is $A \ll 1$, as expected. On the other hand, we compared our prediction, $\Dot{T}$, with $\Dot{P}_b^{corr}$, a corrected version of $\Dot{P}_b$ (kinematic bias) and we found that, in the case of PSR J$1012 + 5307$, GR  fails to explain the observed values, while an additional non-local contribution could explain them, provided that $0.12 < A < 0.16$. This means that such a gravity theory should be further investigated. However, it must be said that the inaccuracies on masses and distances are notable, and vary greatly in the literature. Moreover, since for PSR J$1012 + 5307$ scintillation, i.e.  enhanced pulse intensity variations with relatively short timescales and narrow frequency bandwidths, is suspected, as emerged in a recent census by LOFAR \citep{LOFAR},  corruptions in pulsar timing, due to the interstellar medium, could be hidden.    This could be a possible explanation for the GR disagreement. Therefore, only by using more precise data and a statistically larger number of sources of this type could confirm, modify or exclude the constraints we found.

\section*{Declaration of competing interest}
The author declare that they have no known competing financial interests or personal relationships that could have appeared to
influence the work reported in this paper.

\section*{Data availability}
No data was used for the research described in the article.

\section*{Acknowledgements}
AC acknowledges the Istituto Nazionale di Fisica Nucleare (INFN), Sezione di Napoli, iniziativa specifica QGSKY for
the support.





\bibliographystyle{elsarticle-harv} 
\bibliography{biblio2}

\end{document}